\documentclass{iopart}
 \ifx\pdfoutput\undefined
   \usepackage[dvips]{graphicx}
   \else
   \usepackage[pdftex]{graphicx}
   \fi
   \usepackage{epstopdf}
\usepackage{iopams}  

\begin{document}
\DeclareGraphicsExtensions{.pdf,.gif,.jpg}

\title{Competing $\mathcal{PT}$ potentials and re-entrant $\mathcal{PT}$ symmetric phase: particle in a box}
\author{Yogesh N. Joglekar$^1$ and Bijan Bagchi$^2$}
\address{$^1$Department of Physics, Indiana University Purdue University Indianapolis (IUPUI), Indianapolis, Indiana 46202, USA}
\address{$^2$Department of Applied Mathematics, University of Calcutta, 92, Acharya Prafulla Chandra Road, Kolkata 700 009, India}
\ead{$^1$yojoglek@iupui.edu, $^2$bbagchi123@rediffmail.com}
\begin{abstract}
We investigate the effects of competition between two complex, $\mathcal{PT}$-symmetric potentials on the $\mathcal{PT}$-symmetric phase of a ``particle in a box''. These potentials, given by $V_Z(x)=iZ\mathrm{sign}(x)$ and $V_\xi(x)=i\xi\left[\delta(x-a)-\delta(x+a)\right]$, represent long-range and localized gain/loss regions respectively. We obtain the $\mathcal{PT}$-symmetric phase in the $(Z,\xi)$ plane, and find that for locations $\pm a$ near the edge of the box, the $\mathcal{PT}$-symmetric phase is strengthened by additional losses to the loss region. We also predict that a broken $\mathcal{PT}$-symmetry will be restored by increasing the strength $\xi$ of the  localized potential. By comparing the results for this problem and its lattice counterpart, we show that a robust $\mathcal{PT}$-symmetric phase in the continuum is consistent with the fragile phase on the lattice. Our results demonstrate that systems with multiple, $\mathcal{PT}$-symmetric potentials show unique, unexpected properties. 
\end{abstract}
\section{Introduction}
\label{sec:intro} 
Over the past fifteen years, there has been tremendous progress in the field of parity and time-reversal ($\mathcal{PT}$) symmetric quantum theory~\cite{r0,r1,r2,r3}; this progress is further spurred by recent experiments on optical waveguides~\cite{expt1,expt2,expt25} and electrical circuits~\cite{expt3}. As it was in the case of original quantum theory, the theoretical developments have occurred on two distinct fronts. The first is investigations of Schr\"{o}dinger equation in the presence of non-Hermitian, $\mathcal{PT}$-symmetric, complex potentials $V(x)=V^{*}(-x)\neq V^*(x)$~\cite{bb1,bb2}.  The second is the study of non-Hermitian, $\mathcal{PT}$-symmetric band matrices~\cite{z1,z2,z3} that can represent the Hamiltonian for a tight-binding lattice with on-site, balanced, loss-and-gain potentials~\cite{expt2,expt25}. 

The developments on these two fronts have been with little overlap. The continuum studies have focused on properties of the Schr\"{o}dinger equation in the complex co-ordinate plane~\cite{bendercomplex}; notable early  exceptions are ``particle in a box'' problem with two simplest potentials, $V_Z(x)=iZ\mathrm{sign}(x)$~\cite{zz,bagchi,most} and $V_\xi(x)=i\xi\left[\delta(x-a)-\delta(x+a)\right]$~\cite{zxi}. In either of these cases, there is a nonzero threshold, measured in the units of relevant energy scale, for the emergence of complex eigenvalues. In corresponding lattice models with complex, constant potential or a pair of impurities, the $\mathcal{PT}$-breaking threshold, measured in the units of tunneling, vanishes as lattice size $N$ diverges~\cite{bendix} except in the case of closest or farthest impurities~\cite{song,mark}. However, on both fronts, the $\mathcal{PT}$-symmetric potential is typically characterized by a single parameter and the energy eigenvalues are real as long as the {\it magnitude} of that parameter is smaller than the threshold value. 

In this paper, we study the effect of competition between smooth ($V_Z$) and localized ($V_\xi$) complex potentials on the $\mathcal{PT}$-symmetric phase diagram. We consider a particle in a box of size $2L$ in the presence of $V(x)=V_Z(x)+V_\xi(x)$, and its lattice counterpart, and calculate the phase diagram in the $(Z,\xi)$ plane as a function of the localized-gain position $a$. Surprisingly, we find that a broken $\mathcal{PT}$-symmetric state induced by one potential can be restored by additional losses to the loss region, with corresponding gains in the gain region, introduced by the second potential. We find that the phase diagrams for $a\leq L/2$ and $a>L/2$ are qualitatively different. With scaling analysis, we show that the robust phases of the continuum model are consistent with the fragile phases of its lattice counterpart. 

\section{Particle in a box with $\mathcal{PT}$-symmetric potentials}
\label{sec:pb}
Let us consider a particle of mass $m$ confined to a line $|x|\leq L$ with Dirichlet boundary conditions for its wave function, $\psi(\pm L)=0$. When $Z>0$, the complex potential $V_Z(x)=iZ\mathrm{sign}(x)$ represents a constant gain potential for $x>0$, with a mirror-symmetric loss region for $x<0$. On the other hand, the complex potential $V_\xi(x)$ represents localized gain ($i\xi$) and loss ($-i\xi$) at mirror-symmetric positions $\pm a$ for $\xi>0$. The Schr\"{o}dinger equation for eigenfunctions of the particle is given by 
\begin{equation}
-\frac{\hbar^2}{2m}\partial^2_x\psi_n(x)+ V(x)\psi_n(x)=E_n\psi_n(x).
\end{equation}
It is straightforward to parametrize the eigenfunctions as~\cite{zxi} 
\begin{equation}
\psi_n(x)=\left\{
\begin{array}{cc}
A\sin[k_n(L-x)]  & a\leq x\leq L, \\
B\cos(k_nx) + C k_n^{*} \sin(k_n x) & 0\leq x\leq a,\\
B\cos(k_n^* x) +C k_n\sin(k_n^* x) & -a\leq x\leq 0,\\
F\sin[k_n^*(L+x)] & -L\leq x\leq -a.\\
\end{array}\right.
\end{equation}
This parametrization guarantees that the wave function and its first derivative are continuous at the origin, and that the wave function vanishes at $\pm L$. The effect of $\delta$-functions at $x=\pm a$ on the eigenfunctions is given by the discontinuity in their derivatives at locations $\pm a$,
\begin{equation}
\label{eq:adiscont}
\partial_x\psi_n(\pm a^{+})-\partial_x\psi_n(\pm a^{-}) = \mp i(2m\xi/\hbar^2)\psi_n(\pm a).
\end{equation}
Eq.(\ref{eq:adiscont}) along with the continuity of the wave function at $x=\pm a$ provide four relations among the four unknown coefficients $A, B, C, F$. The complex wave vector $k_n$ 
satisfies $E_n-iZ = \hbar^2k_n^2/2m$~\cite{zxi}. In the following analysis, we choose $\hbar^2/2m=1=L$, use the dimensionless quantities $k_n=k_n L$, $p=a/L$, $Z=Z/(\hbar^2/2mL^2)$, and $\xi=\xi/(\hbar^2/2mL)$ and employ the notation $C_u=\cos(u)$ and $S^{*}_u=\sin(u^*)$ to simplify calculations. The resulting characteristic equation for $k_n$ is 
\begin{eqnarray}
\label{eq:char}
\Re\left[|k_n|^2 k_n C_{k_n} S^*_{k_n} + \xi^2|S_{(1-p)k_n}|^2 k_n C_{p k_n} S^{*}_{pk_n}\right]& + & \nonumber \\ 
 i\xi\Im\left[ |k_n|^2S_{(1-p)k_n}C_{pk_n} S^{*}_{k_n}- k_n^2 S^{*}_{(1-p)k_n} S^{*}_{pk_n} C_{k_n}\right] & = & 0. 
\end{eqnarray}
We characterize the wave vector as $k_n=\pi\alpha_n+i\beta_n$. Simultaneous solutions of Eq.(\ref{eq:char}) and the constraint $-Z=2\pi\alpha\beta$, obtained graphically, lead to the permitted values of $k_n$ and dimensionless energies $E_n=(\pi\alpha)^2-\beta^2$ respectively. As the strength of the complex potential is increased, two adjacent or next-nearest neighbor eigenvalues become degenerate and then presumably complex, leading to the breaking of the $\mathcal{PT}$ symmetry. When $\xi=0$, the threshold potential strength is $Z_{c}\sim 4.48$~\cite{zz} whereas for $Z=0$ and $a=L/2$, it is given by $\xi_c\sim 5.06$~\cite{zxi}. In the two cases, the eigenvalues $E_n$ are real for $-Z_c\leq Z\leq Z_c$ and $-\xi_c\leq \xi\leq\xi_c$ respectively. {\it When both of them are nonzero}, the $\mathcal{PT}$-symmetric phase in the $(Z,\xi)$ plane can be characterized by two boundary functions, $Z_{c-}(\xi)\leq Z\leq Z_{c+}(\xi)$ with $\xi\geq 0$. 

It is straightforward to verify following properties of the characteristic equation: i) When $Z=0$, the term in Eq.(\ref{eq:char}) proportional to $i\xi$ vanishes. ii) When $p\rightarrow 0, 1$ the $\xi$-dependence of Eq.(\ref{eq:char}) vanishes. iii) When $\xi=0$, the eigenvalues $E_n$ are even in $Z$. iv) The $\mathcal{PT}$-phase boundaries for $\xi<0$ are given by $Z_{c+}(\xi)=-Z_{c-}(-\xi)$ and $Z_{c-}(\xi)=-Z_{c+}(-\xi)$. Therefore, we focus only on the region $\xi\geq 0$ in the $(Z,\xi)$ plane; when $Z=0=\xi$, Dirichlet boundary conditions imply that $k_n=\pi\alpha_n+i\beta_n=n\pi/2$. 

\section{$\mathcal{PT}$-phase diagram and the re-entrant $\mathcal{PT}$-symmetric phase}
\label{sec:phase}

\begin{figure}[h!]
\begin{center}
\begin{minipage}{16cm}
\begin{minipage}{8cm}
\hspace{-5mm}
\includegraphics[angle=0,width=7.5cm]{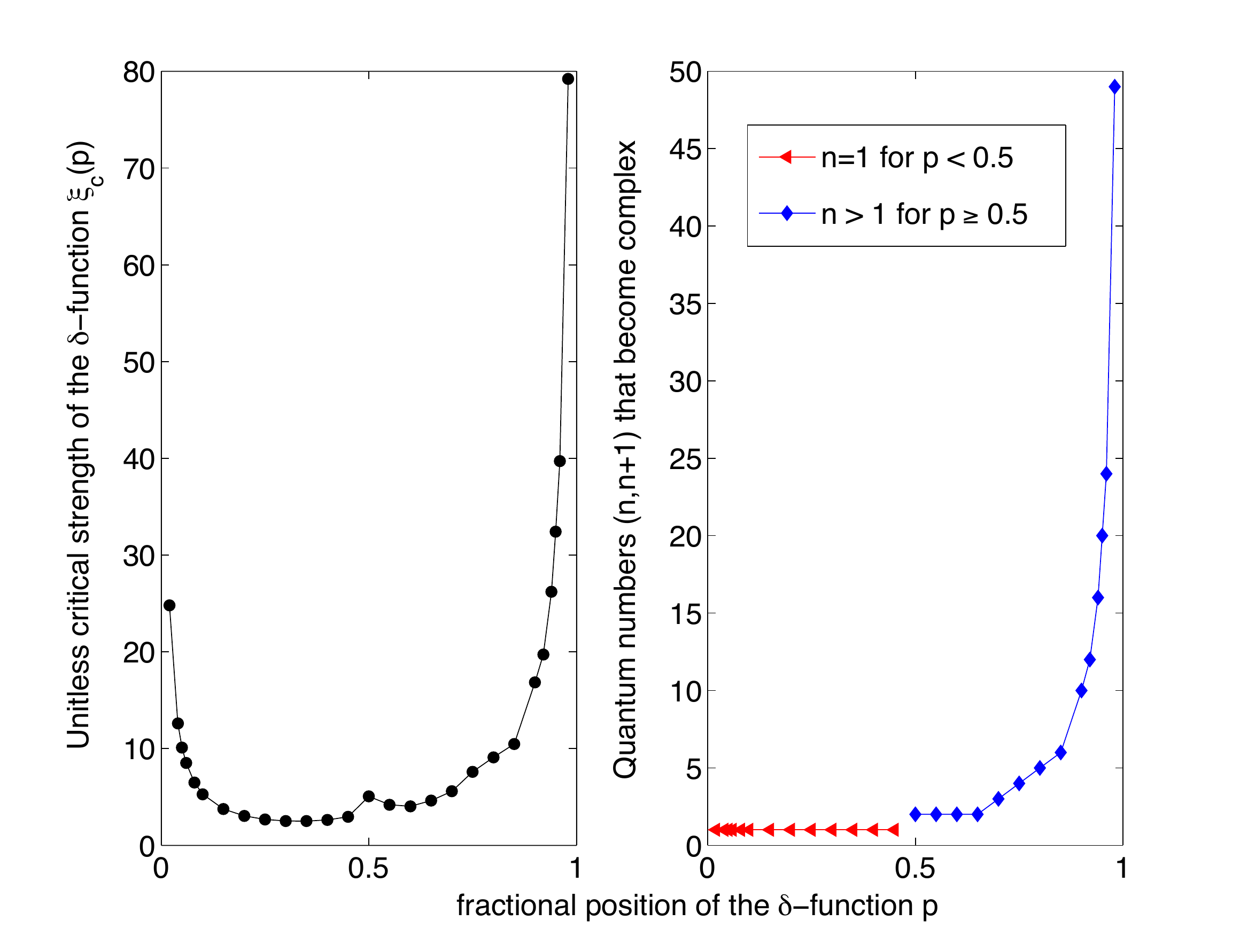}
\end{minipage}
\hspace{-10mm}
\begin{minipage}{8cm}
\includegraphics[angle=0,width=6.5cm]{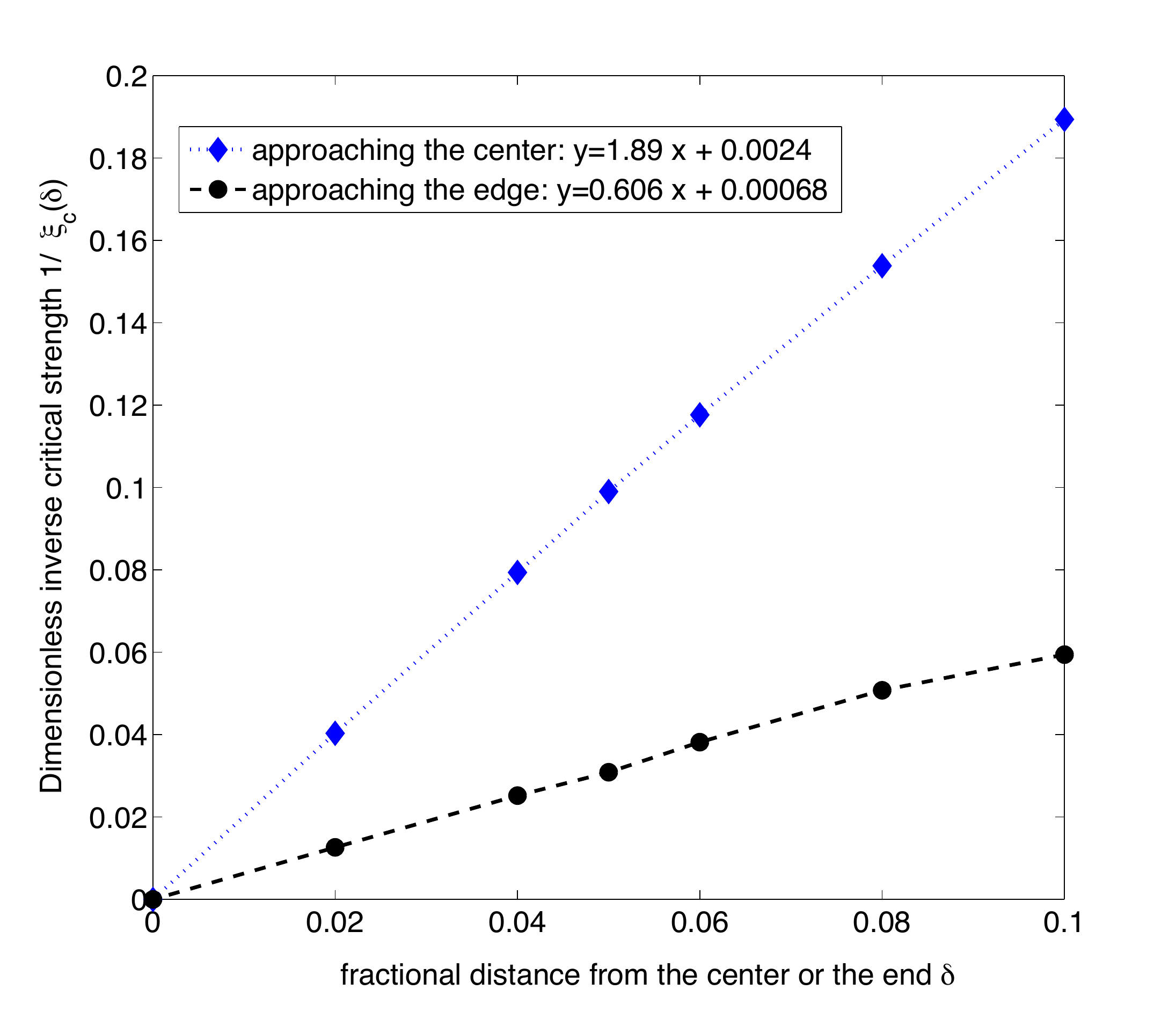}
\end{minipage}
\end{minipage}
\caption{$\mathcal{PT}$-phase diagram for the potential $V_\xi$. The left-hand panel shows the threshold strength $\xi_c(p)$ as a function of the fractional position $p=a/L$ of the localized-gain potential. Apart from a local maximum at $p=0.5$~\cite{zxi}, the threshold $\xi_c(p)$ diverges asymmetrically as $p\rightarrow 0,1$. The center panel shows the quantum number $n$ for which the energies $(E_n,E_{n+1})$ become degenerate at the $\mathcal{PT}$ symmetric threshold. For $p<0.5$ (red triangles), the $\mathcal{PT}$-symmetry breaks at the ground-state energy $n=1$. For $p\geq 1/2$ (blue diamonds), the corresponding index $n(p)$ increases and diverges as $p\rightarrow 1$. The right-hand panel shows the scaling of the diverging threshold strength $\xi_c$ as function of fractional distance $\delta\ll 1$ from the origin ($p=0$, blue diamonds) or the boundary ($p=1$, black circles). The vertical axis denotes $1/\xi_c(\delta)$ and shows that $\xi_c(\delta)\propto\delta^{-1}$.}
\label{fig:xiphase}
\end{center}
\end{figure}
We start this section with results for the $\mathcal{PT}$-symmetric threshold $\xi_c(p)$ when $Z=0$; the properties of this model near $p=0.5$ have been extensively explored~\cite{zxi}. The left-hand panel in Fig.~\ref{fig:xiphase} shows the phase diagram $\xi_c(p)$ as a function of $0<p<1$. Starting from the local maximum at $p=0.5$, as the localized-gain location moves towards the origin, $p\rightarrow 0$, or the boundary, $p\rightarrow 1$, the threshold strength $\xi_c(p)$ first decreases and then increases asymmetrically. The center panel shows the index $n$ for which the eigenvalues $E_n$ and $E_{n+1}$ become degenerate and then presumably complex when the $\mathcal{PT}$ threshold is crossed. For $p<0.5$, $n=1$ (red triangles) whereas for $0.5\leq p<1$, the eigenvalues that become degenerate have increasingly higher energy (blue diamonds).\footnote{Note that for certain rational values of $p$ including $p=0.5$, the eigenvalue $E_{n+1}(\xi)$ is independent of $\xi$, and energies $(E_n, E_{n+2})$ become degenerate leading to $\mathcal{PT}$-symmetry breaking~\cite{zxi}.} The right-hand panel shows the scaling of the threshold strength for impurity locations close to the origin, $\delta=p$, (blue diamonds) or the boundary, $\delta=1-p$ (black circles). For $\delta\ll 1$, we see that threshold diverges as $\xi_c(p)\propto\delta^{-1}$ although the prefactor for the divergence is smaller when $p\rightarrow 0$ (blue diamonds) than when $p\rightarrow 1$ (black circles).  Thus, the $\mathcal{PT}$-symmetric phase diagram at $Z=0$ is acutely sensitive to the distance between the localized gain and loss positions. 

\begin{figure}[tbh]
\begin{center}
\begin{minipage}{18cm}
\begin{minipage}{8cm}
\hspace{-5mm}
\includegraphics[angle=0,width=7cm]{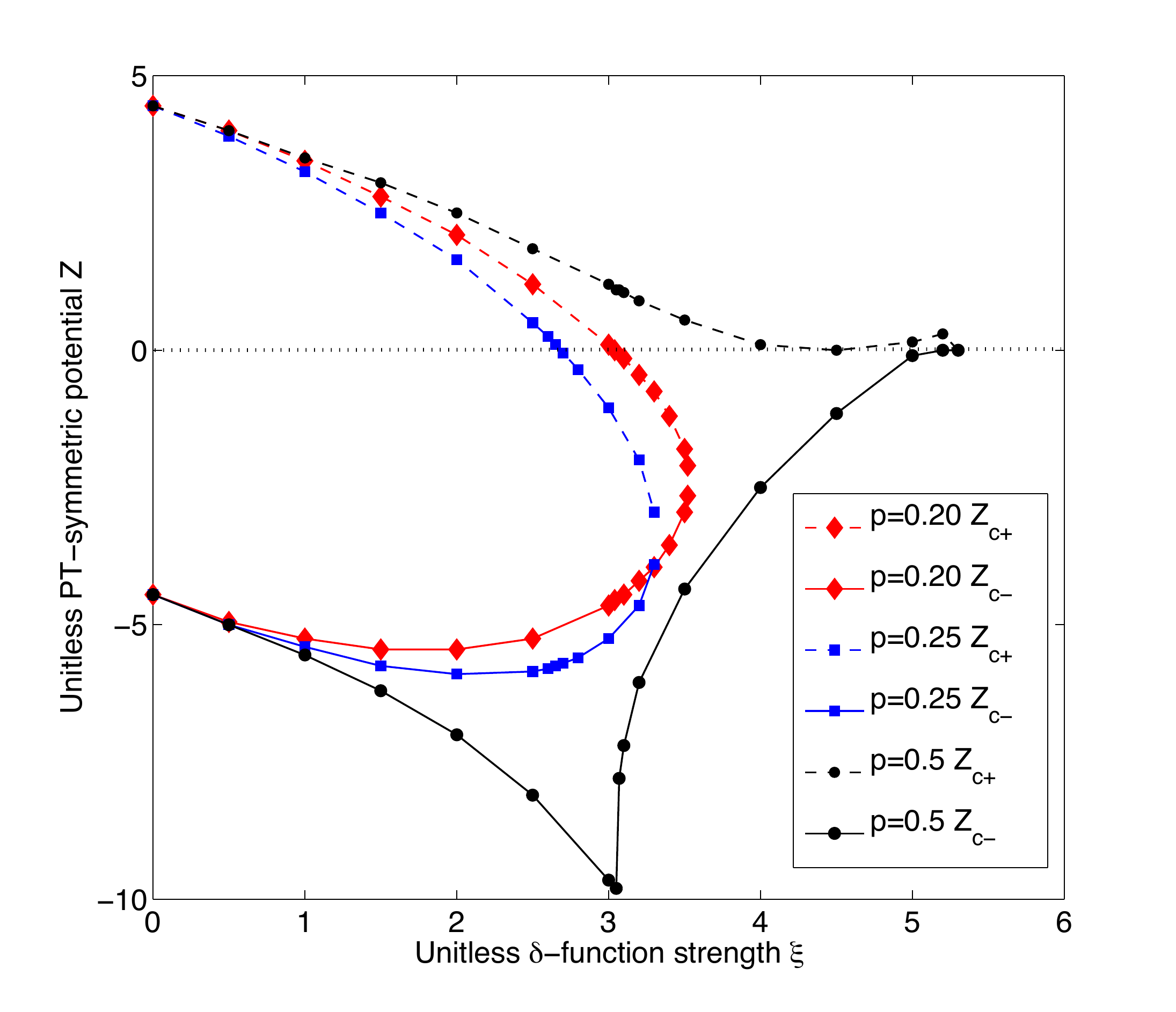}
\end{minipage}
\hspace{-15mm}
\begin{minipage}{8cm}
\includegraphics[angle=0,width=7cm]{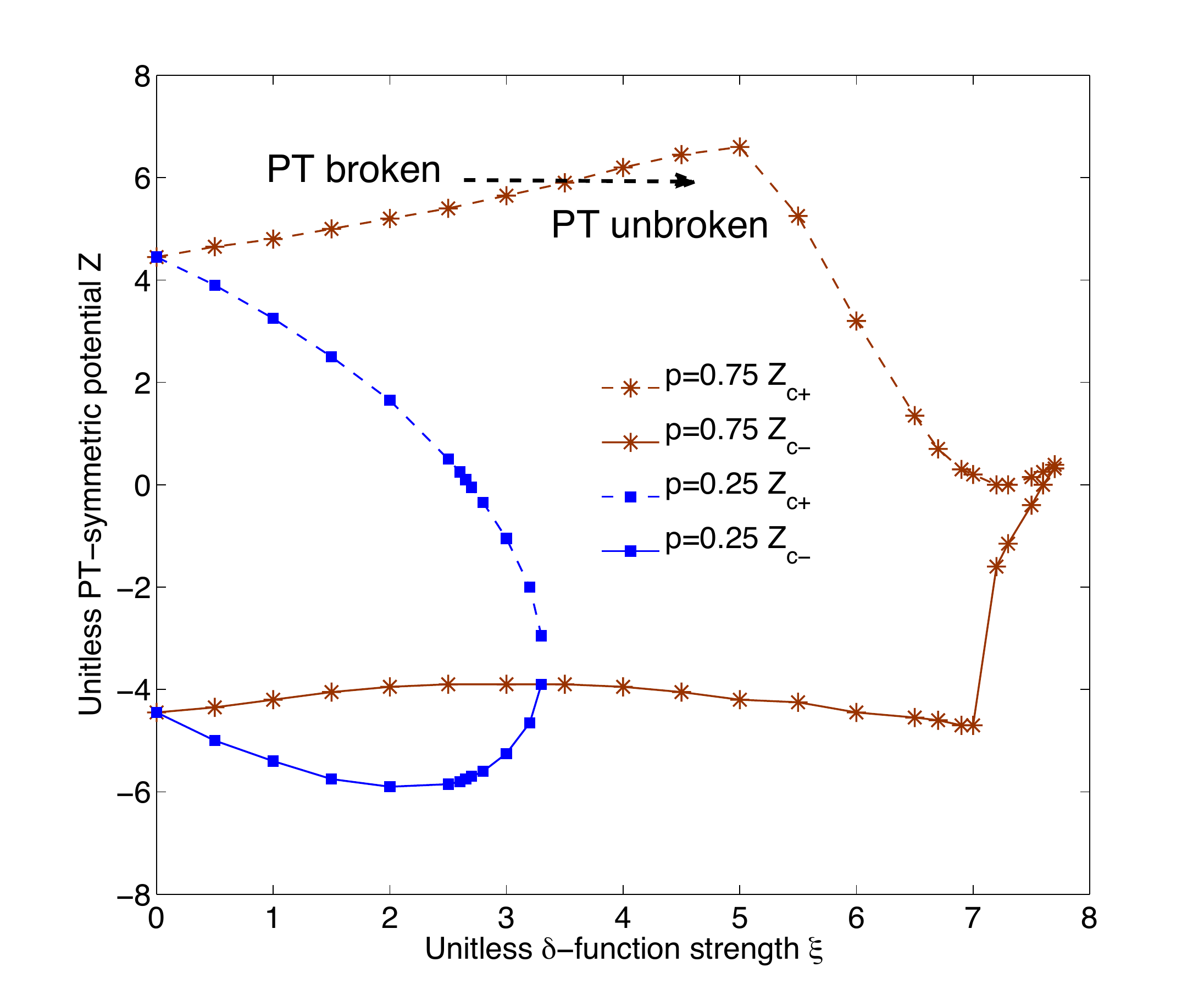}
\end{minipage}
\end{minipage}
\caption{$\mathcal{PT}$-symmetric phase diagram in the $(Z,\xi)$ plane. The left-hand panel shows the $\mathcal{PT}$-phase boundaries $Z_{c\pm}(\xi)$ for $\xi\geq 0$ and different locations $p\leq 0.5$ of the localized-gain impurity. In general, as $\xi$ increases, the threshold strengths $Z_{c\pm}$, shown by dashed and solid lines respectively, decrease. The right-hand panel contrasts the results for $p=0.75$ (brown stars) and $p=0.25$ (blue squares). When $p=0.75$, the threshold strength $Z_{c+}(\xi)$ (brown stars, dashed line) increases with $\xi$ while $Z_{c-}(\xi)$ (brown stars, solid line) remains approximately constant. This surprising result implies that the broken $\mathcal{PT}$ symmetry at $Z\gtrsim 4.5, \xi=0$ is restored by adding localized-gain $\xi\sim 5$ to the constant-gain region (dashed arrow). } 
\label{fig:phasediagram}
\end{center}
\end{figure}
Now we present the phase diagram in the $(Z,\xi)$ plane for $0\leq\xi\leq\xi_c(p)$ as a function of the localized-gain location $p$. The left-hand panel in Fig.~\ref{fig:phasediagram} shows the phase boundaries $Z_{c+}(\xi)$ (dashed lines) and $Z_{c-}(\xi)$ (solid lines) for $p=0.5$ (black circles), $p=0.25$ (blue squares), and $p=0.20$ (red diamonds). When $\xi=0$, the phase boundaries are symmetric about the vertical axis, $Z_{c+}(\xi)=-Z_{c-}(\xi)\sim 4.48$~\cite{zz}. When $\xi>0$, the phase boundaries become asymmetrical about $Z=0$. In general, as the gain strength $\xi>0$ for the localized potential increases, the corresponding thresholds for the constant potential $Z_{c}$ decrease. This behavior is not unexpected. The $\mathcal{PT}$-symmetry is unbroken when the ``effective potential strength", naively proportional to $|\xi+Z|$, is smaller than a threshold value. Therefore, as $\xi$ increases, maintaining the same ``effective strength" requires lowering the threshold values $Z_{c\pm}(\xi)$ including driving them negative. 

As we will show now, this argument is not generally applicable. The right-hand panel in Fig.~\ref{fig:phasediagram} shows the $\mathcal{PT}$-phase boundaries for $p=0.75$ (brown stars), along with results for $p=0.25$ (blue diamonds) for comparison. When $p=0.75$, as the localized gain strength $\xi$ increases from zero, the corresponding gain threshold $Z_{c+}(\xi)$ for the constant potential strength also increases whereas the loss threshold $Z_{c-}(\xi)$ remains essentially constant. It implies that a broken $\mathcal{PT}$-symmetric phase that occurs at, say $Z\gtrsim 4.5$ and $\xi=0$, will be restored by the addition of a localized-gain impurity $\xi\sim 5$ at position $p=0.75$ to the constant-gain region. Thus, contrary to standard expectations, we predict that the $\mathcal{PT}$-symmetric phase is {\it strengthened by adding two gain potentials}. We emphasize that this re-entrant behavior of the $\mathcal{PT}$-symmetric phase is generic and becomes more prominent as $p\rightarrow 1$. This surprising result shows that the two complex, $\mathcal{PT}$-symmetric potentials can lead to a co-operative effect that enhances the $\mathcal{PT}$-symmetric threshold.  

\section{Continuum vs. lattice results}
\label{sec:lattice}
 This continuum model has nonzero thresholds $Z_c$ or $\xi_c(p)$ below which the $\mathcal{PT}$-symmetry is unbroken. On the other hand, the lattice models of these two potentials have a fragile $\mathcal{PT}$-symmetric phase, defined by vanishing threshold strengths, except when the localized gain or loss impurities are closest or the farthest~\cite{bendix,song,mark}. We now focus on the apparent discrepancy between these two results. Let us consider a lattice with $N\gg 1$ sites, nearest-neighbor spacing $a_L$ and tunneling $t_0$, with open boundary conditions. In the continuum limit, $a_L\rightarrow 0$, $N\rightarrow\infty$ such that $2L=Na_L$ remains finite. In addition, when $a_L\rightarrow 0$, the tunneling $t_0\rightarrow\infty$ such that the product $t_0 a_L^2=\hbar^2/2m$ remains finite and gives the ``effective mass" $m$ of the particle hopping on this lattice. 

For the localized gain and loss potentials the $\delta$-function strength $\xi=a_L \gamma$ where $\pm i\gamma$ denotes the on-site impurity potential in the lattice version.  Thus, the continuum and lattice thresholds for fractional position $p$ are related by $\xi(p)/(\hbar^2/2m L)=N\gamma_c(p)/t_0$. For almost all values of $p$, the lattice phase is fragile, $\gamma_c(p)/t_0\propto N^{-1}\rightarrow 0$ as $N\rightarrow\infty$~\cite{mark}. However, due to its precise $N^{-1}$ scaling, the continuum threshold $\xi_c(p)$, measured in units of continuum scale, remains nonzero~\cite{zz}. On the other hand, finite threshold strength $\gamma_c/t_0\sim 1$ for closest or farthest impurities implies that the corresponding continuum threshold diverges as $\xi_c/(\hbar^2/2mL)\sim N$. This is precisely the result shown in the right-hand panel in Fig.~\ref{fig:xiphase}; closest or farthest impurities on a lattice translate into continuum potentials located at a distance $\delta\propto 2/N$ away from the origin or the boundary respectively.

\begin{figure}[h]
\begin{center}
\includegraphics[angle=0,width=7cm]{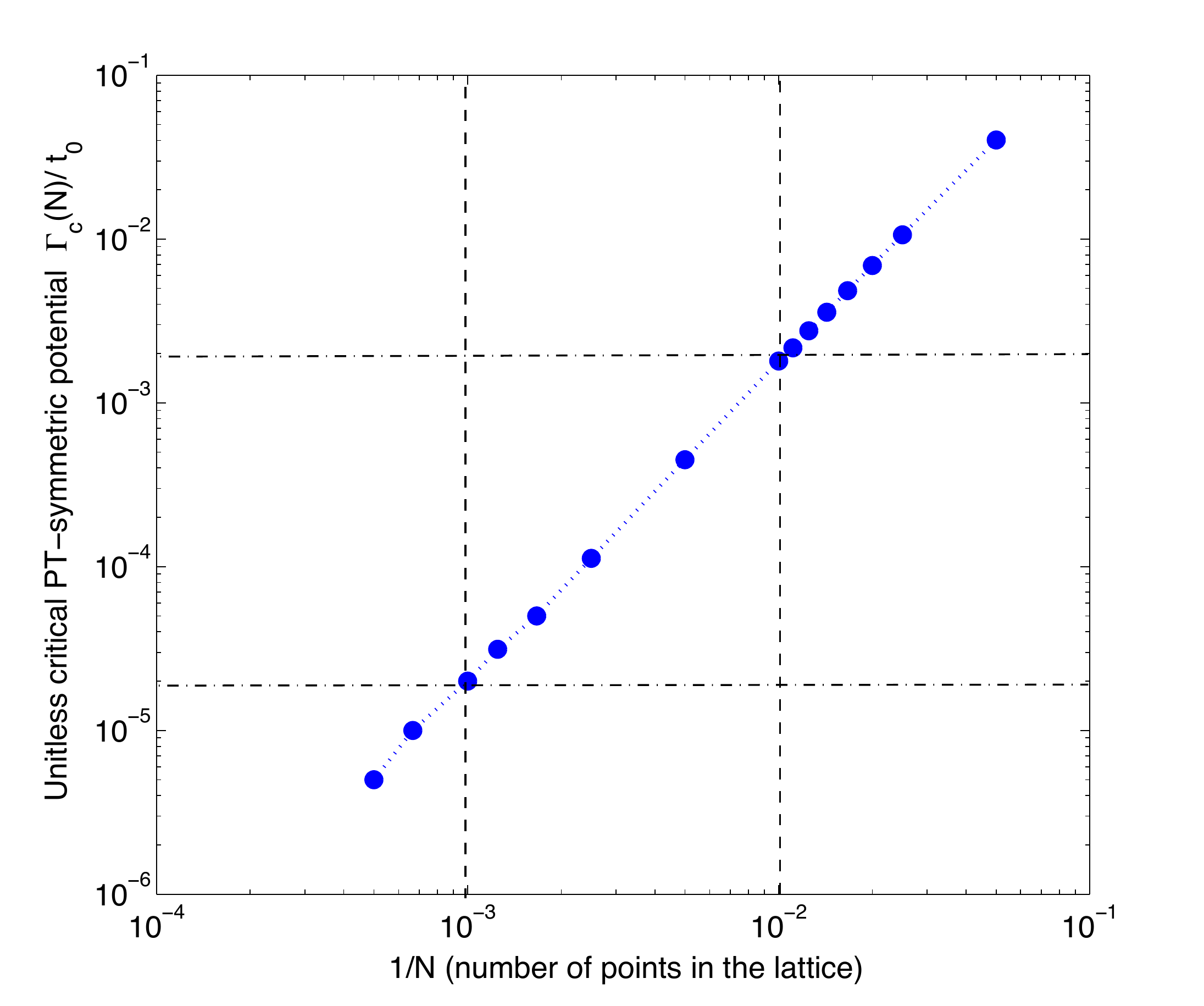}
\caption{$\mathcal{PT}$-symmetry breaking threshold $\Gamma_c(N)/t_0$ as a function of the lattice size $N$ for a tight-binding lattice with tunneling $t_0$ and complex, on-site potential $V_j=i\Gamma\mathrm{sign}(j-N/2)$; note the logarithmic scale on both axes. Robust phase of the corresponding continuum problem~\cite{zz} implies that the corresponding $\mathcal{PT}$-phase on the lattice is fragile with the lattice threshold $\Gamma_c/t_0\propto N^{-2}$; the results for lattices with $N=20$ to $N=2500$ (blue circles) show that it is indeed the case.}
\label{fig:zscaling}
\end{center}
\end{figure}
For the constant gain and loss potential $V_Z$, a corresponding argument shows that $Z_c/(\hbar^2/2m L^2)=N^2\Gamma_c/t_0$ where the lattice-version of $V_Z(x)$ is given by the on-site potential $V_j= i\Gamma\mathrm{sign}(j-N/2)$ for site index $j=1,\ldots,N$, and $\Gamma_c$ is the critical impurity strength for the lattice Hamiltonian. Since the continuum threshold is finite $Z_c/(\hbar^2/2mL^2)\sim 4.48$, we expect that the corresponding $\mathcal{PT}$-symmetric phase on the lattice will be fragile with scaling behavior $\Gamma_c/t_0\propto N^{-2}$. Figure~\ref{fig:zscaling} shows the critical threshold $\Gamma_c(N)/t_0$ for lattice sizes $N=20$ to $N=2500$; note the logarithmic scale on both axes. The dotted horizontal and vertical lines span one and two decades respectively, and thus the numerically obtained results (blue circles) show excellent agreement with the prediction that follows from continuum-limit considerations. 


\section{Conclusions}
\label{sec:conc}

In this paper, we have investigated the competition between two complex, $\mathcal{PT}$-symmetric potentials on the $\mathcal{PT}$-symmetric phase diagram. We have shown that a (simple) 
system with such potentials supports an unusual phase diagram and the restoration of $\mathcal{PT}$-symmetric phase. At present, the most promising or current candidates for experimental realization of a $\mathcal{PT}$-system can be modeled by tight-binding lattices and are explored via time-evolution of an initial state $|\phi\rangle$ that is localized to a few, if not single, lattice sites~\cite{expt2,expt25,expt3}. Such a localized state contains energy components across the entire lattice bandwidth $4t_0$, and therefore can only probe the fragile $\mathcal{PT}$-symmetric phase of the lattice. 

We have shown that the robust or fragile nature of the $\mathcal{PT}$-symmetric phase is essentially determined by the relevant, continuum or lattice, energy scales. Thus $\mathcal{PT}$-symmetry breaking effects in the continuum problem, including re-emergence of the $\mathcal{PT}$-symmetric phase with increasing potential strength, can be probed through the dynamics of a broad and smooth initial state $|\Phi\rangle$ with size $1\ll M\ll N$ that contains only the energy components near the bottom of the energy band. An experimental investigation of such a system will deepen our understanding of $\mathcal{PT}$-symmetry breaking in continuum and lattice models.


\section{acknowledgment}
\label{sec:ack}
BB acknowledges the warm hospitality of the Physics department at IUPUI where this work was initiated. This work is supported, in part, by NSF grant DMR-1054020. 




\section*{References}
\begin{thebibliography}{99}
\bibitem{r0} Bender C M and Boettcher S 1998 {\it Phys. Rev. Lett.} {\bf 80} 5243
\bibitem{r1}Bender C M, Brody D C, and Jones H F 2002 {\it Phys. Rev. Lett.} {\bf 89} 270401
\bibitem{r2} For a review, see Bender C M 2007 {\it Rep. Prog. Phys.} {\bf 70} 947
\bibitem{r3} For a review, see Mostafazadeh A 2010 {\it Int. J. Geom. Meth. Mod. Phys.} {\bf 7} 1191 (arXiv:08105643)
\bibitem{expt1} Guo A {\it et al.} 2009 {\it Phys. Rev. Lett.} {\bf 103} 093902
\bibitem{expt2} R\"{u}ter C E, Makris K G, El-Ganainy R, Christodoulides D N, Segev M, and Kip D 2010 {\it Nat. Phys.} {\bf 6} 192
\bibitem{expt25} Feng L {\it et al.}, 2011 {\it Science} {\bf 333} 729
\bibitem{expt3} Schindler J, Li A, Zheng M C, Ellis F M, and Kottos T 2011 {\it Phys. Rev. A} {\bf 84} 040101(R)
\bibitem{bb1} Cavaglia A, Fring A, and Bagchi B 2011 {\it J. Phys. A} {\bf 44} 325201
\bibitem{bb2} Bagchi B and Quesne C 2010 {\it J. Phys. A} {\bf 43} 305301
\bibitem{z1} Znojil M 2012 {\it Phys. Lett. A} {\bf 375} 3435
\bibitem{z2} Znojil M 2010 {\it Phys. Rev. A} {\bf 82} 052113
\bibitem{z3} Znojil M 2007 {\it J. Phys. A} {\bf 40} 13131
\bibitem{bendercomplex} Bender C M 2010 {\it Phys. Rev. Lett.} {\bf 104} 061601
\bibitem{zz} Znojil M 2001 {\it Phys. Lett. A} {\bf 285} 7
\bibitem{bagchi} Bagchi B, Mallik S., and Quesne C. 2002 {\it Mod. Phys. Lett. A} {\bf 17} 1651
\bibitem{most} Mostafazadeh A and Batal A 2004 {\it J. Phys. A} {\bf 37} 11645
\bibitem{zxi} Znojil M and Jakubsk\'{y} V 2005 {\it J. Phys. A} {\bf 38} 5041
\bibitem{bendix} Bendix O, Fleischmann R, Kottos T, and Shapiro B 2009 {\it Phys. Rev. Lett.} {\bf 103} 030402
\bibitem{song} Jin L and Song Z 2009 {\it Phys. Rev. A} {\bf 80} 052107
\bibitem{mark} Joglekar Y N, Scott D D, Babbey M, and Saxena A 2010 {\it Phys. Rev. A} {\bf 82} 030103(R)
\end{thebibliography}
\end{document}